\documentstyle[12pt]{article}
\begin{document}
\baselineskip=15pt
\def\bzeta{\mbox{\boldmath $\rho $}}
%--------------------------------------------------------------------------
\renewcommand{\baselinestretch}{1.0}
\renewcommand{\theequation}{\arabic{equation}}
\input epsf
%----------------------------------------------------------------------------
\title{\bf{Elastic moduli renormalization in self
interacting stretchable polyelectrolytes}} 
\author{ Rudi Podgornik$\dagger$$\thanks{To whom
correspondence should be addressed. E-mail address:
rudi@helix.nih.gov. }$, Per Lyngs Hansen and V. Adrian
Parsegian 
\\{ } 
\\{Laboratory of Structural Biology} 
\\{National Institute of Child Health and Human Development}
\\{National Institutes of Health, Bethesda, MD 20892-5626}
\\{and}
\\{$\dagger$ Department of Physics, Faculty of Mathematics and Physics}
\\{University of Ljubljana, SI-1000 Ljubljana, Slovenia }
\\{and Department of Theoretical Physics}
\\{J. Stefan Institute, SI-1000 Ljubljana, Slovenia}}

\begin{titlepage}
\maketitle
\begin{abstract}
\small
%-------------------------------------------------------------------------
\noindent 

We study the effect of intersegment interactions on the effective
bending and stretching moduli of a semiflexible polymer chain with a
finite stretching modulus.  For an interaction potential of a screened
Debye-H\" uckel type renormalization of the stretching modulus is
derived on the same level of approximation as the celebrated
Odijk-Skolnick-Fixman result for the bending modulus.  The presence of
mesoscopic intersegment interaction potentials couples the bending and
stretching moduli in a manner different from that predicted by the
macroscopic elasticity theory.  We advocate a fundamental change in
the perspective regarding the dependence of elastic moduli of a
flexible polyelectrolyte on the ionic conditions: stretchability.  Not
only are the persistence length as well as the stretching modulus
dependent on the salt conditions in the solution, they are
fundamentally coupled via the mesoscopic intersegment interaction
potential.  The theory presented here compares favorably with recent
experiments on DNA bending and stretching.

%-------------------------------------------------------------------------
\end{abstract}
\end{titlepage}
%-------------------------------------------------------------------------

\section{Introduction}

Mechanical properties as characterised by bending, stretching and
twisting and their respective elastic moduli, more than anything else
determine the supercoiling \cite{dna} and packing properties of DNA
\cite{book}.  The effect of intersegment interactions along a
semiflexible polymer chain on its bending modulus, especially in the
context of DNA, has been studied for quite a while (see Ref. 
\cite{manning} and references therein).  The major result of these
studies is the celebrated Odijk-Skolnick-Fixman (OSF) formula
\cite{osf} that connects the value of the persistence length with the
parameters of the interaction potential between the segments along the
polymer chain usually assumed to be of the Debye-H\" uckel form.  

In the case of a stretchable chain there exists no theory that would
connect the intersegment interactions and the stretching elastic
modulus.  That this theory is very much needed is shown by recent
experiments on single molecule DNA stretching \cite{Baumann}.  They
point to the conclusion that the measured renormalization of the
stretching modulus in the presence of solutions with different ionic
strengths can not be rationalized on the basis of simple elastic
theory arguments \cite{LL} according to which the renormalization of
the bending and stretching moduli should be proportional.  Experiments
on the contrary suggest that while the presence of electrostatic
repulsion between DNA segments tends to stiffent up the chain it also
makes it a lot more stretchable.  This would suggest that DNA might
not be describable by macroscopic elasticity theory at all
\cite{Baumann}.

The physical basis of the bending modulus renormalization is quite
simple and well understood, see Fig.  \ref{fig0}.  The change in the
persistence length is due to the fact that the effective spacing
between the segments gets smaller ($L' < L$) after the polymer is
locally bent.  Because the intersegment interactions are assumed to be
repulsive the interactions oppose bending and thus give rise to a
higher bending modulus.  The physics of the effect of the intersegment
interactions on the stretching modulus, being proportional to the
second derivative of the interaction energy as a function of the
intersegment coordinate, is quite different.  If the chain is allowed
to stretch locally the segment length becomes bigger ($a' > a$), the
interaction energy becomes less steep and its second derivative thus
becomes smaller.  Therefore the more the segments are further appart
the less the second derivative of the interaction energy is going to
be, leading to a smaller stretching modulus.  In what folllows we will
formalize and extend this simple physical picture.

One should note here that part of this effect already transpires
through the work of Ha and Thirumalai \cite{Ha}.  Though they deal
with a nominally unstretchable chain, the unstretchability coinstraint
is implemented globally through an appropriate Lagrange multiplier. 
In the presence of intersegment interactions this Lagrange multiplier
would be renormalized by the interaction.  Our calculation builts on
and adds to this change in the Lagrange multiplier by explicitely
introducing a stretching part of the elastic energy.

In this article we present a straightforward generalization of the OSF
arguments to include the effect of the intersegment interactions on
the stretching modulus of the chain as well.  We derive both the
bending modulus or equivlanetly the persistence length renormalization
as well as the stretching modulus renormalization concurrently on the
same level of approximation based on the recent implementation of the
$1/d$ expansion technique into the self-interacting semiflexible
polymer theory \cite{Hansen}.  We show that for finite range
intersegment interaction potentials the bending and stretching
renormalization become coupled.  Comparison with recent experiments
seems to bear out our line of thinking quite strongly. 

We are convinced that our calculations dispell any doubts as to 
whether DNA conforms to macroscopic elasticity theory \cite{Baumann}. 
It does, if one takes the long range part of the interaction potential 
between the segments of the polymer chain consistently into account.

The outline of the paper is as follows: First we introduce a
mesoscopic model of the self-interacting polymer chain with bending
and stretching elasticity included.  We briefly describe the 1/d
expansion method that we use to evaluate the partition function of the
model. We then explicitely obtain a mean field solution of the model 
and the contribution of the fluctuations to the equation of state of 
the polymer chain that connect the elongation of the chain with the 
stretching force acting on it. We finally derive the renormalized 
elastic moduli of the chain and compare the results with available 
experiments.

\section{Model}

We start by formulating an elastic mesoscopic Hamiltonian for a self
interacting chain described in the highly stretched, small deformation
limit in the Monge-like parameterization (see Fig.  \ref{fig1}) as
${\bf r}(s) = (z, \bzeta (z))$, where $s$ is the arclength along the
chain.  For a one-dimensional solid \cite{LL} which is an adequate
representation of a flexible polymer chain on small lengthscales, the
deformation tensor has only one non-zero component
\begin{equation}
u_{zz}(z) = {{\partial u_z(z)}\over{\partial z}} + {\textstyle{1\over 2}}
\left( {{\partial \bzeta(z)}\over{\partial z}}\right)^2,
\label{two}
\end{equation}
where $u_z(z)$ is the internal phonon-like field describing the
stretching of the chain.  The bending field $\bzeta(z)$ is in the
direction perpendicular to the local tangent of the chain, thus
perpendicular to the $z$ axis.  This result can be derived
straightforwardly from the form of the line element along the chain:
$ds^2 = (dz + du_z(z))^2 + (d\bzeta(z))^2$ \cite{LL}.

The total mesoscopic energy of a semi-flexible self interacting chain
under external tension contains three terms: the stretching elastic
energy term, the bending elastic energy term, the stretching force term and
the interaction term.  It can be written straightforwardly in the form
\begin{eqnarray}
{\cal H} &=& {\textstyle {1 \over 2}}\lambda \int \left( {{\partial
u_z(z)}\over{\partial z}}dz + {\textstyle{1\over 2}} \left( {{\partial
\bzeta(z)}\over{\partial z}}\right)^2\right)^2 dz + {\textstyle {1
\over 2}} K_C \int \left( {{\partial^2 \bzeta(z)}\over{\partial z^2}}
\right)^2 dz - \nonumber\\ &-& f \int {{\partial u_z(z)}\over{\partial
z}}dz + {\textstyle {1 \over 2}} \int\int V(\vert {\bf r}(z) - {\bf
r}(z') \vert ) dzdz',
\label{three}
\end{eqnarray}
where $V(\vert {\bf r}(z) - {\bf r}(z') \vert )$ is the interaction
potential between two segments of unit length separated by $\vert {\bf
r}(z) - {\bf r}(z') \vert^2 = \left( z - z' + u_z(z) - u_z(z')
\right)^2 + \left( \bzeta(z) - \bzeta(z') \right)^2 $. $f$ is the
external force stretching the chain in direction $z$, $\lambda$ is the
stretching (Lam\' e) modulus and $K_C$ is the bending modulus related
to the persistence length $l_P$ as $K_C = kT~l_P$. This mesoscopic 
energy presents a generalization of the existing models of a stretched 
elastic chain, Refs. \cite{Ha}, \cite{marko} and \cite{Odijk}.

\section{The 1/d Expansion Method}

The non-local nature of the intersegment interaction, dependent on
both $z$ and $z'$, precludes an explicit evaluation of the partition
function of the model with a mesoscopic Hamiltonian Eq.  \ref{three}. 
We thus have to resort to some approximation scheme that will make
the evaluation of equilibrium properties of this model tractable.

At this point we introduce the 1/d expansion method to obtain an
approximate but nevertheless explicit form of the partition function. 
The basis of this method is the introduction of two auxiliary fields:
$B(z,z') = \left( {\bf r}(z) - {\bf r}(z')\right)^2$ and its Lagrange
multiplier $g(z,z')$ (for details see \cite{Hansen}) that will help
transform non-local intersegment interactions along the chain into
local energy terms.  With these variables, and limiting ourselves to at
most quadratic order in all the variables, the chain mesoscopic
Hamiltonian can be cast into the form
\begin{eqnarray}
{\cal H} &=& {\textstyle {1 \over 2}}\lambda \int \left( {{\partial
u_z(z)}\over{\partial z}} \right)^2 dz + 
{\textstyle {1\over 2}} K_C \int \left( {{\partial^2 \bzeta(z)}\over{\partial z^2}}
\right)^2 dz -  f \left( u_z(L) - u_z(0)\right) + \nonumber\\
&+& {\textstyle {1 \over 2}} \int\int dzdz' V(B(z,z')) + \nonumber\\
&+& {\textstyle {1 \over 2}} \int\int dzdz'~ g(z,z')\left( B(z,z') - \left( z - z' + u_z(z) - u_z(z')
\right)^2 + \left( \bzeta(z) - \bzeta(z') \right)^2\right) , \nonumber\\
.
\label{four}
\end{eqnarray}
where we indicated explicitely the dependence of the interaction
potential on the auxiliary field ${B(z,z')}$ as $V(B(z,z')) =
V(\sqrt{B(z,z')})$.  This dependence is introduced via the Lagrange
multipier in the last line of the above equation Eq.\ref{four}
through the constraint $B(z,z') = \left( {\bf r}(z) - {\bf
r}(z')\right)^2$.

The rationale for this change of variables is that the dimensions of
the fields $\bzeta(z)$ and $u_z(z)$, that can be integrated over
explicitely and exactly, are assumed to be much larger than the
dimensions of the auxiliary fields ${B(z,z')}$ and $g(z,z')$.  This
allows the contribution of the auxiliary fields to the partition
function to be evaluated on the saddle-point level.  This approach can
be shown to be asymptotically exact if the dimension of the embedding
space for the polymer chain, {\sl i.e.} the dimension of the ${\bf
r}(s)$ vector, tends to infinity.  If this is not the case, as in deed
it is not for our three dimensional case, what we get is a result
valid to ${\cal O}(1/d)$.  Even in this case the $1/d$ method gives
reasonable results that compare very favorably with other methods (see
Ref.  \cite{Hansen} and the references cited therein).

The free energy is now obtained by the standard trace over the
fluctuating fields as well as the auxiliary fields and their Lagrangian
multipliers
\begin{eqnarray}
{\cal F} &=& - kT~\ln{\int\dots\int {\cal D}u(z){\cal
D}{\bzeta}(z){\cal D}g(z,z'){\cal D}B(z,z') e^{- \beta {\cal H}} }.
\label{five}
\end{eqnarray}
Once we have an explicit form for the free energy we can get an
equation of state for the self-interacting chain, connecting the 
external stretching force acting on the chain with its elongation, from
\begin{eqnarray}
     - \frac{\partial {\cal F}}{\partial f} = \left< \left( u_z(L) -
     u_z(0)\right)\right> = (L - L_{0}),
     \label{eqstate}
\end{eqnarray}    
where $L$ is the length of the chain after and $L_{0}$ before the onset
of the external stretching force.

\section{The Mean Field Solution}

Before integrating over all the fluctuating fields let us investigate
the mean-field solutions of the mesoscopic Hamiltonian Eq. 
\ref{four}.  Let us first imagine we have no external tension applied
to the chain, {\sl i.e.} $f = 0$.  The presence of the intersegment
interactions, however, acts as an effective stretching force by itself. 
Let us see how that happens.  The mean-field solution for this case is
obtained by minimizing the Hamiltonian Eq.  \ref{four} and assuming
that all the fields are constant:
\begin{equation}
u_{zz} =  \zeta , ~~{\rm and}~~\bzeta(z) = 0.
\end{equation}
Thus since
\begin{eqnarray}
{\cal H}_{0} &=& {\textstyle {1 \over 2}}\lambda \int dz \zeta^2 +
{\textstyle {1 \over 2}} \int\int dzdz' V(B(z,z')) + \nonumber\\
&+& {\textstyle {1 \over 2}} \int\int dzdz' g(z,z') \left( B(z,z') -
\left( z - z' \right)^2\left( 1 + \zeta\right)^2 \right),
\end{eqnarray}
by minimizing with respect to $\zeta$ we obtain the equation of state in the form
\begin{equation}
    \zeta = \frac{\delta \lambda}{\lambda - \delta \lambda},
    \label{hooke}
\end{equation}
where we introduced
\begin{equation}
    \delta\lambda =  \int dz' g(z,z') \left( z - z'\right)^2.
    \label{trap}
\end{equation}
The mean field minimization with respect to the auxiliary fields gives
\begin{eqnarray}
    B(z,z') &=& \left( z - z' \right)^2\left( 1 + \zeta\right)^2 
    \nonumber \\
    g(z,z') &=& - \partial_{B}V(B(z,z')).
    \label{clap}
\end{eqnarray}    
We note at this point that the above mean field equations are highly 
and essentially non-linear. First of all $g(z,z')$ is a non-linear 
functional of $B(z,z')$, Eq. \ref{clap}, and $ \delta\lambda$ is 
determined from a solution to Eq. \ref{trap}.

Thus on this level we see that the intersegment interactions stretch
the chain in a way similar to an external force, leading to what one
could call a Hooke's law of the form Eq.  \ref{hooke}.  If we now add
a real external tension ($f$) to the chain the mean-field ansatz would
assume the form
\begin{equation}
u_{zz} =  \zeta + \delta \zeta(f), ~~{\rm and}~~\bzeta(z) = 0.
\end{equation}
The corresponding Hamiltonian in this case is
\begin{eqnarray}
{\cal H}_{0} &=& {\textstyle {1 \over 2}}\lambda \int dz (\zeta + 
\delta\zeta)^2 - f \int dz \delta\zeta +
{\textstyle {1 \over 2}} \int\int dzdz' V(B(z,z')) + \nonumber\\
&+& {\textstyle {1 \over 2}} \int\int dzdz' g(z,z') \left( B(z,z') -
\left( z - z' \right)^2\left( 1 + \zeta + \delta\zeta\right)^2 \right).
\label{mf}
\end{eqnarray}
Obviously we have coupled the stretching force only to the deformation
($\delta\zeta$) after the intrinsic deformation ($\zeta$) set by the
intersegment interactions has been already established.

Minimizing with respect to $\delta \zeta$ we now get
\begin{equation}
    \delta\zeta = \frac{f}{\lambda - \delta \lambda},
\end{equation}
and minimization with respect to the other variables gives
\begin{eqnarray}
    B(z,z') &=& \left( z - z' \right)^2\left( 1 + \zeta + \delta\zeta\right)^2 
    \nonumber \\
    g(z,z') &=& - \partial_{B}V(B(z,z')).
    \label{be}
\end{eqnarray}    
Putting the two results together, the mean-field theory thus gives for
the total deformation
\begin{equation}
    \zeta + \delta\zeta = \frac{f + \delta \lambda}{\lambda - \delta \lambda}.
\label{kra}
\end{equation}
Again it is quite obvious that the intersegment interactions act in a
way similar to an additional stretching force.  Since the intersegment
interactions make an additive contribution to $f$, see Eq.  \ref{kra},
they obviously just displace the mean field minimum around which the
system fluctuates.

\section{Fluctuations}

Now that we have the mean field solution for the case with external
stretching force as well as intersegment interactions, we can expand
the mesoscopic Hamiltonian around the mean field and evaluate also the
effect of thermal fluctuations.  To second order this expansion yields
\begin{eqnarray}
    {\cal H} &=& {\cal H}_{0} + {\textstyle {1 \over 2}}\lambda\left(
    \zeta + \delta\zeta \right) \int dz \left( {{\partial 
    \bzeta(z)}\over{\partial z}}\right)^2 + {\textstyle {1 \over 2}}\lambda \int 
    \left( {{\partial u_z(z)}\over{\partial z}} \right)^2 dz +\nonumber \\
    &+&  
    {\textstyle {1\over 2}} K_C \int \left( {{\partial^2 \bzeta(z)}\over{\partial z^2}}
\right)^2 dz - {\textstyle {1 \over 2}} \int\int dzdz' g(z,z') \left( \left( u_z(z) - u_z(z')
\right)^2 + \left( \bzeta(z) - \bzeta(z') \right)^2\right), \nonumber \\
.
\end{eqnarray}
where ${\cal H}_{0}$ is given by Eq.  \ref{mf}.  In order to evaluate
the functional integral corresponding to this effective Hamiltonian we
first of all develop $u_z(z)$ and $\bzeta(z)$ in the last term of the
above equation into a Taylor series with an argument $z-z'$.  This
means that all the properties of the chain are homogeneous and depend
only on $z-z'$ \cite{Hansen}.  We are thus trying to account for the
longest length scale effects of the interaction terms on the
properties of the semiflexible chain (an equivalent procedure would be
to look at the lowest wave vector dependence of the Hamiltonian in the
Fourier space).  Both devices are consistent with a macroscopic
character of the approach advocated here.  The result of this
expansion is as follows:
\begin{equation}
    {\textstyle {1 \over 2}} \int\int dzdz' g(z,z') \left( u_z(z) - 
    u_z(z')\right)^2 = {\textstyle {1 \over 2}} \delta\lambda 
    \int dz \left( {{\partial u_z(z)}\over{\partial z}} \right)^2 + 
    \dots,
\end{equation}
and
\begin{equation}
    {\textstyle {1 \over 2}} \int\int dzdz' g(z,z') \left( \bzeta(z) - 
    \bzeta(z') \right)^2 = {\textstyle {1 \over 2}} \delta\lambda 
    \int dz \left( {{\partial\bzeta(z)}\over{\partial z}}\right)^2 - {\textstyle 
    {1 \over 2}}  \delta K_C \int dz \left( {{\partial^2 
    \bzeta(z)}\over{\partial z^2}}\right)^2  + \dots
\end{equation}
where the dots stand for higher derivative terms and we introduced
\begin{equation}
    \delta K_C = {\textstyle {1 \over 12}}  \int dz' g(z,z') \left( z - z'\right)^4.
\end{equation}
The complete Hamiltonian, including the mean field part as well as the
contribution of fluctuations around the mean field now becomes
\begin{eqnarray}
    {\cal H} &=& {\textstyle {1 \over 2}}\lambda \int dz \left( 
    \zeta + \delta\zeta \right)^2 - f \int dz \delta\zeta +
     {\textstyle {1 \over 2}} \int\int dzdz' V(B(z,z')) + \nonumber \\
      &+& {\textstyle {1 \over 2}}\lambda^{(R)} \int 
    \left( {{\partial u_z(z)}\over{\partial z}} \right)^2 dz + 
    {\textstyle {1 \over 2}} \tilde{f}\int dz 
    \left( {{\partial \bzeta(z)}\over{\partial z}}\right)^2 + {\textstyle 
    {1 \over 2}}  \delta K_C^{(R)} \int dz \left( 
    {{\partial^2 \bzeta(z)}\over{\partial z^2}}\right)^2,
\end{eqnarray}    
where we introduced the following renormalized elastic constants and 
a rescaled stretching force
\begin{eqnarray}
    \lambda^{(R)} &=& \lambda - \delta\lambda \nonumber\\
    K_C^{(R)} &=& K_C + \delta K_C \nonumber\\
    \tilde{f} &=&  \lambda\left( \zeta + \delta\zeta \right) - 
    \delta\lambda =\frac{f + \frac{\delta\lambda^2}{\lambda}}{1 - 
    \frac{\delta\lambda}{\lambda}}.
    \label{renor}
\end{eqnarray}
This functional integral can be evaluated exactly for the harmonic
variables $u_z(z)$ and ${\bzeta}(z)$ assuming that we can ignore the
end effects.  The evaluation of the functional integral over
non-harmonic degress of freedom , {\sl i.e.} for the two auxiliary
fields $B(z-z')$ and $g(z-z')$, is dealt with on the saddle point
level which constitutes the 1/d approximation (for details see Ref. 
\cite{Hansen}) and leads to Eqs.  \ref{be}.

The free energy of the chain can therefore be obtained in the form
\begin{eqnarray}
    {\cal F} &=& -~kT~\ln{\int\dots\int {\cal D}u(z){\cal D}
    {\bzeta}(z){\cal D}g(z,z'){\cal D}B(z,z') e^{- \beta {\cal H}} } = \nonumber\\
    &=& {\textstyle {1 \over 2}}\lambda \int dz \left( 
    \zeta + \delta\zeta \right)^2 + f \int dz \delta\zeta +
     {\textstyle {1 \over 2}} \int\int dzdz' V(B(z,z')) + \nonumber \\
      &+& {\textstyle {{kT }\over 2}} ~\ln{{\rm det}\left(\lambda^{(R)}
      {{\partial^2 }\over{\partial z^2}} \right)} + kT~\ln{{\rm det}\left( K_C^{(R)}{{\partial^4 }\over{\partial z^4}} -
      \tilde{f}{{\partial^2 }\over{\partial z^2}}\right)}.
      \label{free}
\end{eqnarray}    
The fluctuation determinants can be evaluated in the Fourier space by
the standard methods \cite{Felsager}.  Since we have derived an
explicit form for the free energy we can thus obtain the equation of
state from Eq.  \ref{eqstate} in the form
 \begin{equation}
     \xi = \frac{L}{L_{0}} = 1 -
     \frac{kT}{2}\frac{\lambda}{\lambda^{(R)}\sqrt{K_C^{(R)}\tilde{f}}}
     + \frac{f}{\lambda^{(R)}}.
     \label{final}
 \end{equation}    
Obviously the second term on the r.h.s. of the above equation comes
from the transverse ($\bzeta(z)$) fluctuations and is thus entropic in
origin while the last term is the mean field stretching term.  We
could also call them entropic and enthalpic stretching terms.
 
We see immediately that in the case of no intersegment interactions or
if the range of these interactions goes to zero (both of these cases
leading to $\delta\lambda = 0$, see below) the above equation of state
reduces exactly to the one obtained by Odijk \cite{Odijk} and Ha and
Thirumalai \cite{Ha}.  A similar equation of state has also been
obtained by Marko and Siggia \cite{marko} for the case of a chain with
intersegment interactions except that the stretching part was added in
by hand and that the bending and stretching moduli renormalization
were not coupled as they are in Eq.  \ref{final}.
 
\section{Elastic Moduli Renormalization}

We now assume that in a uni-univalent salt solution the intersegment
interaction potential is purely repulsive and of a screened Debye-H\"
uckel form, {\sl i.e.} 
\begin{equation}
    V({\bf r}(z), {\bf r}(z')) = {{kT l_B}\over{a^2}} \frac{\exp{(-\kappa
|{\bf r}(z)- {\bf r}(z')| )}}{| {\bf r}(z)- {\bf r}(z')| }, 
\label{shoo}
\end{equation}
where $l_B$ is the Bjerrum length, $a$ is the effective separation
between the charges along the chain and $\kappa$ is the inverse Debye
length.  With this intersegment potential and assuming the mean field
form for $B(z)$ Eq.  \ref{be} one gets for the interaction driven
changes in the stretching and bending moduli the following relations
\begin{eqnarray}
\delta\lambda = - \int ds~s^2~V'(B(s)) = {{kT l_B}\over{\Delta^3 a^2}} ( 1 -
Ei(-\kappa a)) \nonumber\\ 
\delta K_C = - {\textstyle {1 \over 12}}
\int ds~s^4~V'(B(s)) = {{kT l_B}\over{4 \Delta^4 (\kappa a)^2}},
\label{goo}
\end{eqnarray}
where $Ei(x)$ is the standard integral exponent function and we introduced the
local stretching parameter $\Delta$ as
\begin{equation}
    \Delta^2 = \frac{B(z,z')}{(z-z')^2} = \frac{\left< \left({\bf r}(z) - {\bf 
    r}(z')\right)^2 \right>}{(z-z')^2} = (1 + \zeta + \delta\zeta)^2 = 
    \left(\frac{\lambda + f}{\lambda^{(R)}}\right)^2.
\end{equation}    
Obviously the renormalizations of the elastic moduli depend on the
magnitude of the intersegment interactions (described by $a$) as well
as on their range (set by the Debye length $\kappa^{-1}$).

If the chain is inextensible, $\lambda \longrightarrow \infty$, then
$\Delta \longrightarrow 1$ , the renormalization of the stretching
modulus becomes irrelevant and the renormalization of the bending
modulus (second equation in Eqs.  \ref{goo}) becomes exactly the
Odijk-Skolnick-Fixman result \cite{osf}, as in deed it should.  Also
one realizes that experimentally \cite{Baumann} $\lambda \gg f$ and
thus one usually has
\begin{equation}
    \Delta \sim \lambda / (\lambda - \delta\lambda) \geq 1.
\end{equation}    
The above relation Eq.  \ref{goo} can be thus viewed as a
generalization of the OSF result for the bending as well as stretching
moduli.  Since in the presence of the intersegment interactions
$\Delta$ is a function of $\delta\lambda$ we have a very non-linear
system of equations to solve. The solution would give us 
simultaneously the renormalization of the bending as well as 
the stretching moduli.

Without even solving this set of equations we already know that the
intersegment repulsions renormalize the bending and stretching moduli
in the opposite directions, see Eq.  \ref{renor}.  While the bending
modulus increases, the stretching modulus decreases.  The simple
physical reasons for this were already outlined in the Introduction. 
This resolves completely the conundrum observed in experimental
studies of DNA stretching and bending elasticity \cite{Baumann}.

One can furthermore examine the relation between renormalized, in
effect measured, bending and stretching moduli in more detail.  For
their {\sl bare} values we should have from the standard macroscopic theory
of elasticity \cite{LL} the result
\begin{equation}
K_C = {\textstyle {1\over 4}} \lambda R^2, 
\label{ela}
\end{equation}
where $R$ is the radius of the molecule (for DNA $R$ is between
10~\AA~and 4~\AA~, corresponding to phosphate and major groove radii). 
Obviously for the renormalized values $K_C^{\tiny (R)}$ and
$\lambda^{\tiny (R)}$ this relation does not hold anymore.  Instead we
obtain the following relation between renormalized stretching and
renormalized bending moduli
\begin{equation}
K_C^{\tiny (R)} = {\textstyle {1\over 4}} \lambda^{\tiny (R)} R^2 + {{kT 
l_B}\over{4 \Delta^4 (\kappa a)^2}} \left(  1 + 2(\kappa R)^2 \Delta( 1 -
Ei(-\kappa a))\right).
\end{equation}
The relation Eq.\ref{ela} is thus valid only asymptotically as the
range and/or the magnitude of interactions becomes very small.  Any
polyelectrolyte in the range of conditions where Eq.  \ref{shoo} is
valid should thus behave as a classical macroscopic cylinder if we
take the coupled bending-stretching moduli renormalization due to the
intersegment interactions properly into account.

\section{Comparison with Experiment}

We can now fit the expressions Eq.\ref{goo} to the recent experiments
by Baumann {\sl et al.} \cite{Baumann} where they measure the
simultaneous dependence of the bending as well as stretching moduli of
DNA on the ionic strength in a uni-uni valent electrolyte.  The
renormalized moduli for different values of the added uni-uni valent
electrolyte are extracted from the fit of the experimental entropic
and stretched regimes to the equation of state Eq.  \ref{final},
just below the overstretching transition.

Fitting the dependence of the renormalized bending and stretching
moduli on the inverse Debye screening length to Eqs. \ref{goo},
we can obtain both the bare bending and stretching moduli as well as
the inverse line charge density $a$. Unbiased fits give for the bare
elastic modulus $2.1\times 10^4~pN~{\textstyle \AA}^2$ (corresponding
to $l_P = 511~{\textstyle \AA}$) and the bare stretching modulus
$\lambda = 1511~pN$. The values of the two bare moduli are
completely consistent with Eq. \ref{ela} considering the fact that for
DNA the radius $R$ lies somewhere between the outer phosphate radius
($\sim 10~{\textstyle \AA}$) and the inner radius of the grooves
($\sim 4~{\textstyle \AA}$).

From the fit to Eq.\ref{goo} we also obtain a consistent estimate (in
the sense that it should fit the stretching as well as the bending
modulus data) for $a \sim 2 ~{\textstyle \AA}$.  This estimate is not
particularly accurate because of the large scatter present in the
data, see Fig.  \ref{fig2}, and because of the number of the fitting
parameters.  The experimental scatter is probably due to the fact that
the regime between entropic and enthalpic stretching is quite narrow
and a reliable estimate for the stretching modulus which can only come
from this regime is thus difficult to obtain.  Experiments are currently
under way to gather a much more accurate set of data for the two
elastic moduli \cite{vic}.

\section{Discussion}

The theory presented above giving the coupling between stretching
and bending moduli renormalization in the presence of finite range
intersegment interactions seems to work reasonably when compared to
experiments.  There is of course no {\sl a priori} reason for this and
considering all the approximations that underpin the main results,
Eqs.  \ref{goo}, is perhaps quite surprising.

First of all the result Eq.  \ref{goo} is formaly valid only in the
limit of either a very stiff chain or very large external tension.  If
the stiffness or the tension are finite, we know \cite{Hansen} that
the extended configuration of the chain lying at the bottom of the
Monge-like parameterization is unstable against thermal fluctuations. 
Nevertheless the OSF limit apepars to be stable \cite{inprep} even in
the regime of vanishing stretching and we expect (without any proof at
this point) that the renormalizations Eqs.  \ref{goo} will remain
likewise.  It would neverteless be appropriate to derive a more
sophisticated theory, somewhat along the lines of \cite{Ha} but without
the unstretchability constraint, that would be able to describe the
equation of state for a stretchable self interacting semiflexible
polymer chain for the whole range of stretching forces.

When calculating the fluctuation contribution to the free energy we 
assumed that $B(z,z')$ and $g(z,z')$ are still given by their mean 
field expressions, Eqs. \ref{be}. There is in deed a fluctuation 
contribution to the auxiliary fields but it is in general small and 
would not  fundamentally change the results derived above. It would 
however make the numerics more cumbersome. 

The numbers extracted for the effective charge density along DNA bear
no resemblance to the Manning theory where the value appropriate for
the effective spacing between the charges is $a \sim 7~\AA$. However 
the fits to the experimental data are not particulary stiff and we 
could certainly push the extracted numbers towards this value if we 
wanted. But considering the large scatter in the experimental data 
especially for the stretching modulus we produced an unbiased fit and 
intend to refine it as better data become available. 

Also we note that the form of the equation of state Eq.  \ref{final}
coincides with the one used to extract the values for the elastic
constants \cite{Baumann} only in the no-intersegment-interactions
limit.  In general it gives corrections to this limit dependent on the
salt concentration.  These corrections should mostly affect the
bending modulus which is obtained from the fit to the $\sqrt{f}$
dependent (entropic) part of the equation of state, and only
marginally the stretching modulus, which is obtained from the fit to
the linear part of the equation of state.  This is another source that
could eventually change the fitted parameters.

Apart from all these numerical shortcomings and problems we advocate a
fundamental change in the perspective regarding the dependence of the
elastic moduli of DNA on the ionic conditions.  Not only are the
persistence length as well as the stretching modulus dependent on the
salt conditions in the solution, they are fundamentally coupled.  This
is a consequence of the fact that as soon as the intersegment
interaction potential is of finite range the stretching and bending
themselves become coupled.  This is most clearly exemplified by the
exact form of the equation of state Eq.  \ref{final}.  We believe that
future work on the elasticity of DNA and similar (bio)polymers will
have to take this fact into account.

\section{Acknowledgement}

One of the authors (RP) would like to thank Prof. V. Bloomfield and Dr. 
I. Rouzina for valuable discussions.

\vfill
\eject
%----------------------------------------------------------

\vfill
\eject
%----------------------------------------------------------
\newpage

\section{\bf Figure captions:}

\vspace{1pt}
\noindent
{\bf Fig. 1} The physics of elastic moduli renormalization in the presence
of finite range intersegment interactions.  Repulsive intersegment
interactions make the chain more difficult to bend because of
deminished effective separation between neighboring segments ($L' <
L$).  However they also make stretching easier because they incerase
the average length of the segments and thus diminish the curvature of
the interaction energy of the segments (symbolically depicted as
beads).

\vspace{10pt}
\noindent
{\bf Fig. 2} A highly stretched polymer chain. The average direction of
the chain is along $z$ axis, which corresponds also to the stretching
axis and is set by the direction of external force $f$ acting at both
ends, and the bending deformation is perpendicular to it. 

\vspace{10pt}
\noindent
{\bf Fig. 3} Experimental points taken from Table 1.  of Baumann {\sl et
al.} \cite{Baumann} and fits using the functional form of the ionic
strength dependence from Eq.\ref{goo}.  $\circ$ are the measurements
of persistence length ( left scale), $\bullet$ are the measurements of
the stretch modulus (right scale) for ionic strengths between 1.86 and
586 mM. The functional form of the two moduli seems to fit the data
quite well.

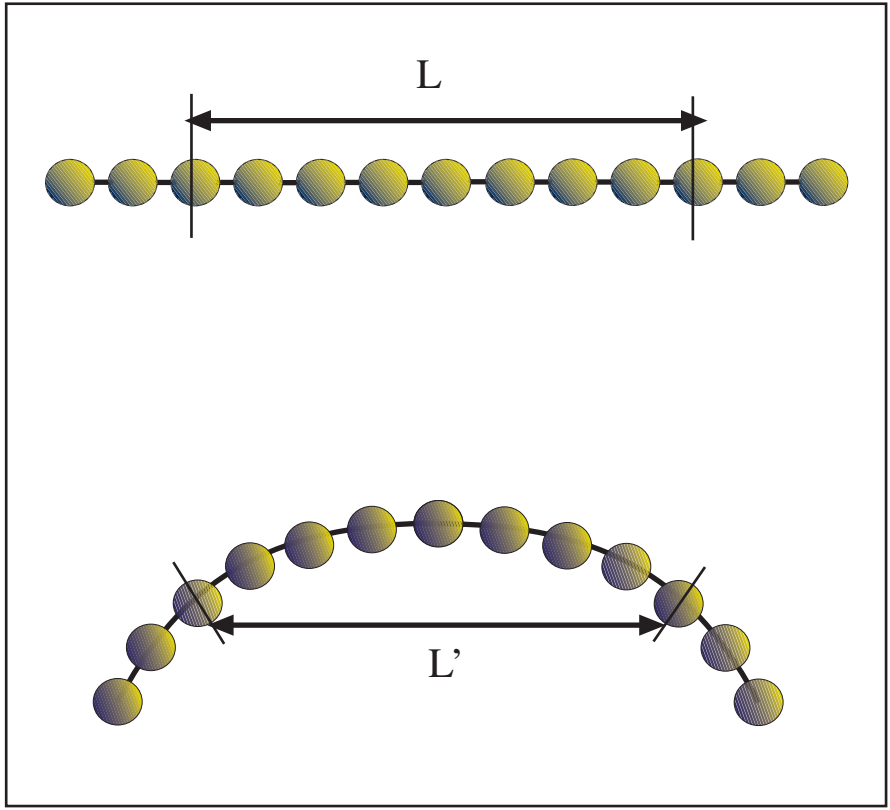
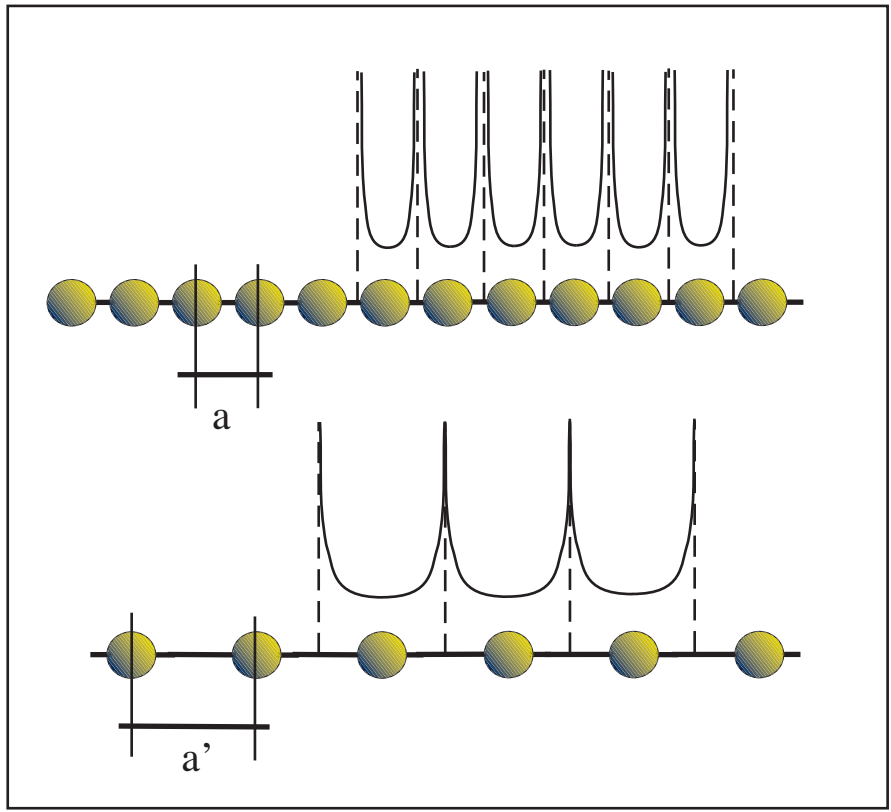
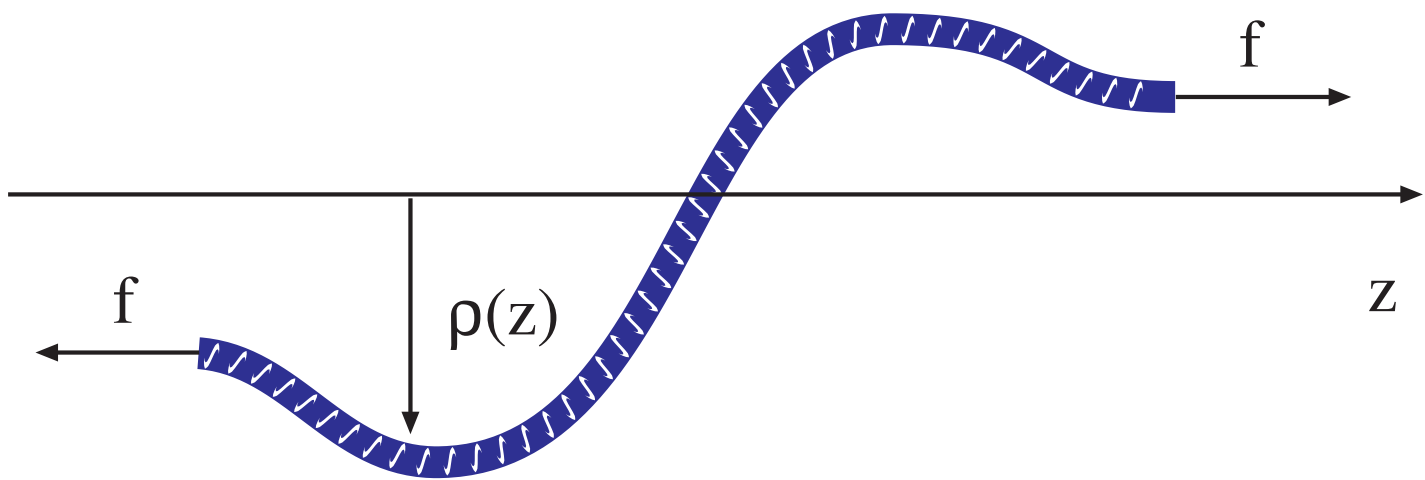
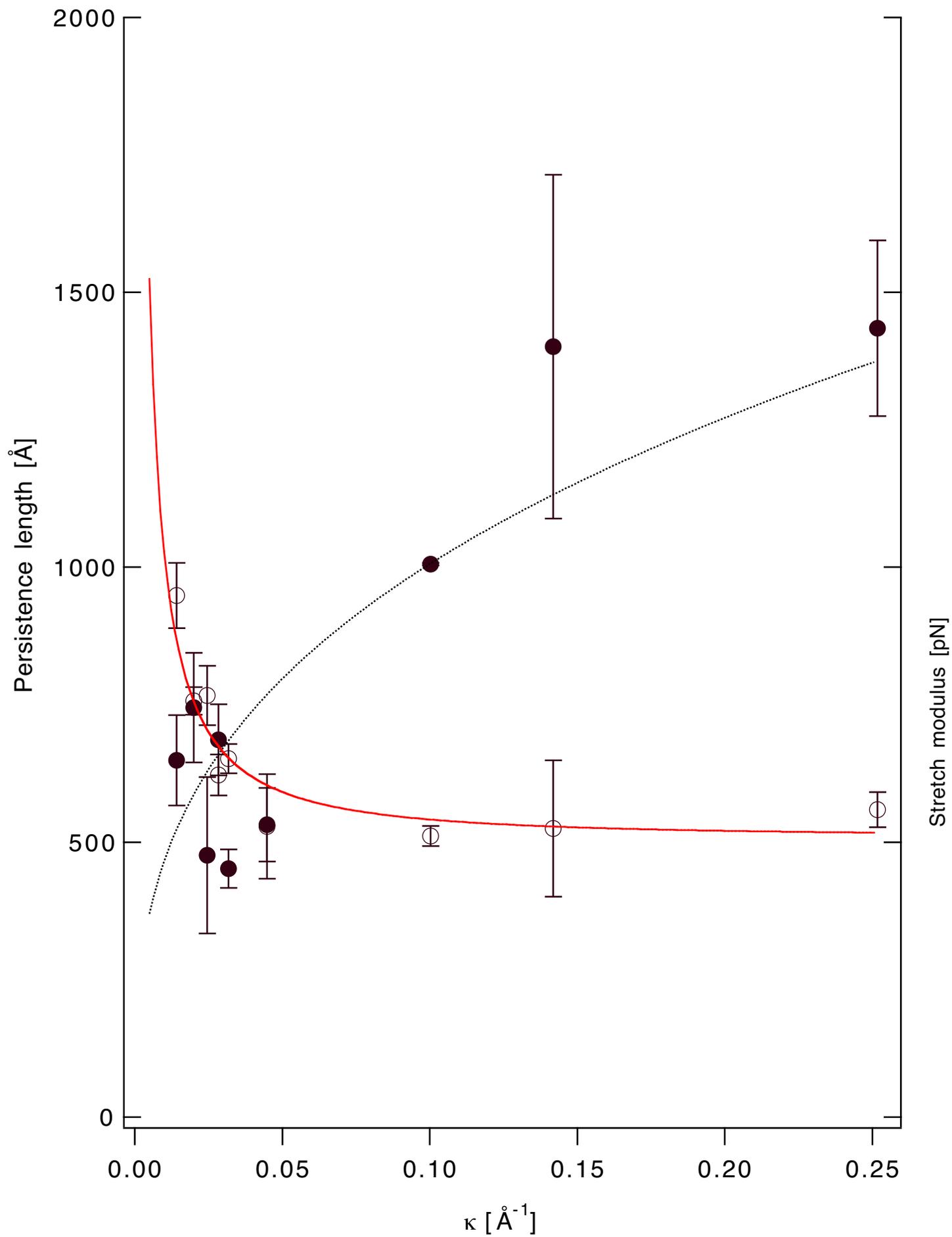
\end{document}